\documentclass[12pt,twoside]{article}

\input{psfig}

\newlength{\dinwidth}
\newlength{\dinmargin}
\newcommand{\beq}[1]{\begin{equation}\label{#1}}
\newcommand{\eeq}{\end{equation}}
 
\newcommand{\beqar}[1]{\begin{eqnarray}\label{#1}}
\newcommand{\eeqar}{\end{eqnarray}}

\newcommand{\gev}{GeV$^2$}
 
\newcommand{\Gl}[1]{Eq.~(\ref{#1})}
\newcommand{\Ab}[1]{Fig.~\ref{#1}}
\newcommand{\Ta}[1]{Table~\ref{#1}}

\newcommand{\al}{\alpha}
\newcommand{\be}{\beta}

\newcommand{\De}{\Delta}

\newcommand{\La}{\Lambda}

\newcommand{\Si}{\Sigma}

\newcommand{\as}{\alpha_s}

\begin{document}

\begin{flushright}
\tt{NIKHEF/97-029} 
\end{flushright}
\vspace{1cm}
\begin{center}
{\Large \bf 
Future measurements of {\boldmath $\alpha_s$}
       and {\boldmath $xg$} from scaling violations at 
       HERA}\\
\vspace{0.8cm}
M.A.J. Botje\footnote{Talk presented
at the 5th International Workshop on Deep Inelastic Scattering
and QCD,
Chicago, IL, April 14--18, 1997}\\
NIKHEF,
PO Box 41882, 1009DB Amsterdam, the Netherlands\\
\vspace{1cm}
\end{center}

\begin{abstract}
\noindent
Results are presented of a study of the experimental and
theoretical accuracy one may achieve at HERA in measuring the strong
coupling constant $\as$ and the gluon distribution
from scaling violations of $F_2$
structure functions.
\end{abstract}

\section*{\centering I\lowercase{ntroduction}}

Accurate measurements of $F_2$ structure functions in deep
inelastic scattering provide one of the cleanest tests of
perturbative QCD. In the first few years of experimentation
at HERA the available kinematic range was extended to low
values of $x \simeq 10^{-4}$ and large $Q^2 \simeq 5000$ \gev.
In the next 8 years of operation HERA might deliver integrated
luminosities of 0.5--1 fb$^{-1}$. The extended kinematic coverage
and increased luminosity allows for detailed measurements of
the scaling violations in $F_2$ and hence of the strong coupling
constant $\as$ and the gluon distribution $xg(x)$ at low $x$.

In this report I will summarise the results of
studies done during the workshop `Future Physics at
Hera'~\cite{mba:hwsc,mba:hwse,mba:hwst} to estimate the 
experimental
and theoretical errors
 on $\as(M_Z^2)$ obtained from a QCD analysis
of future $F_2$ structure function data. 
Also shown is the precision one may reach in the determination
of the gluon distribution.

\section*{\centering C\lowercase{omparison of \uppercase{NLO} evolution codes}}

In this section I present results of a study on
how well the various implementations of perturbative QCD
are numerically and conceptually
under control~\cite{mba:hwsc}.

The evolution equations for the parton distributions $f(x,Q^2)$ in
the proton are given by
\beq{mba:eq1}
\frac{\partial f(x,Q^2)}{\partial \ln Q^2} =
\left[ a_s(Q^2)P_0(x) + a_s^2(Q^2)P_1(x) + O(a_s^3) \right] \otimes f(x,Q^2)
\eeq
where we write $a_s(Q^2) \equiv \alpha_s(Q^2)/4\pi$ and where
$P_0$ and $P_1$ are the leading order (LO) and NLO splitting functions
respectively.
%The scale dependence of the strong coupling constant reads
%\beq{mba:eq2}
%\frac{\partial a_s}{\partial \ln Q^2} = -\be_0 a_s^2(Q^2)
%-\be_1 a_s^3(Q^2) + O(a_s^4)
%\eeq

Two methods are widely used to solve \Gl{mba:eq1}. In the first
approach (`$x$--space') the parton distributions are numerically evolved
on a grid in $x$ and $Q^2$. This method is conceptually simple but
the numerical accuracy depends on the number of gridpoints in $x$ 
and $Q^2$ which is limited by the amount of CPU time one can afford
to spend in the computations. 

In the second
approach (`$N$--space') the Mellin transform of \Gl{mba:eq1} is taken
so that
the convolution integrals become simple products. The resulting
ordinary differential equations can be solved analytically. The result
is then transformed back to $x$--space. This method is mathematically
more involved but accuracies of $\sim 10^{-5}$ are
readily achieved~\cite{mba:th}.

\begin{figure}[b!] % fig 1
\centerline{
\psfig{file=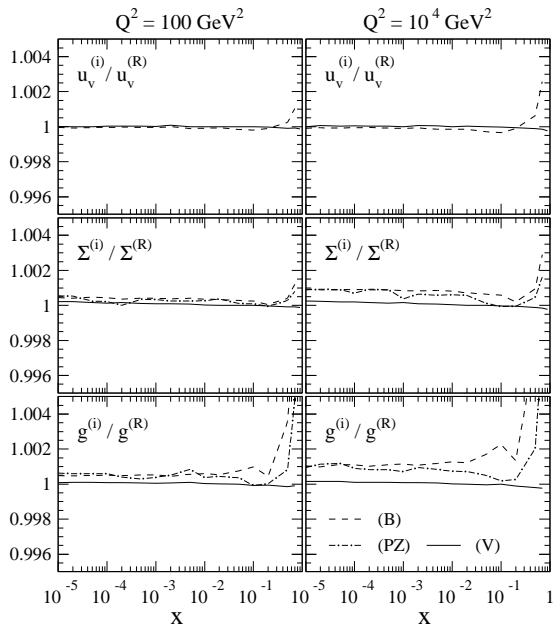,bbllx=20pt,bblly=120pt,bburx=580pt,bbury=700pt,height=9.5cm,clip=}}
\caption{The relative difference between the up-valence ($u_v$), singlet
($\Si$) and gluon ($g$) densities as obtained from evolving identical
input at $Q^2 = 4$ \gev\ in NLO with the evolution programs
$i$ = (B, PZ, V) and (R).}
\label{mba:fig1}
\end{figure}

Detailed comparisons were made of two 
$x$--space programs used by the ZEUS~(B)~\cite{mba:qcdnum}
and H1~(PZ)~\cite{mba:pznum} collaborations and two
$N$--space progams
which we label (V)~\cite{mba:vnum} and (R)~\cite{mba:rnum}
respectively. 
With these four programs identical sets of input
parton distributions ($10^{-5} < x < 1$) were evolved 
in NLO from the
input scale $Q^2 = 4$ \gev\ up to $Q^2 = 10^4$ \gev. 

\Ab{mba:fig1}
shows a comparison of the results. One notices the very good agreement
between the $N$--space programs (V) and (R). The two $x$--space programs
(B) and (PZ) agree to within 0.05\% with the $N$--space programs over
a wide kinematic range. The agreement is slighly worse ($\sim 1\%$)
at very high $x$ where the parton distributions vanish.
We remark that the agreement between the $N$--space and $x$--space
programs can in principle be further improved by increasing the number
of gridpoints in the latter.

\section*{\centering E\lowercase{xperimental errors on \boldmath{$\alpha_s$}}}

To investigate the experimental error on $\as (M_Z^2)$ QCD fits were
performed to  simulated HERA $F_2$ datasets  listed in \Ta{mba:tab1}. 
The data cover a kinematic range 
$1.5 \times 10^{-5} < x < 0.7$ and $0.5 < Q^2 < 5 \times 10^4$ \gev.
\begin{table}[b]
\label{mba:tab1}
\begin{center}
\begin{tabular}{clccccc}
  Dataset & nucleon & $E_{e}$ (GeV) & $E_{N}$  (GeV) &
 $L$ (pb$^{-1}$) &  $Q^{2}_{min}$  (\gev) & $Q^{2}_{max}$ (\gev) \\
\hline     
I  & proton & 27.6   & 820  & 10  & 0.5  & 100   \\
II  & proton & 27.6   & 820  &  500  & 100  & 50000   \\
III & proton & 27.6   & 400  & 200  & 100  & 20000   \\
IV & proton & 15.0   & 820  & 10  & 0.5  & 100   \\
V  & deuteron & 27.6   & 410  & 10   & 0.5  & 100   \\
VI & deuteron & 27.6   & 410  & 50   & 100  & 20000   \\
\hline     
\end{tabular}
\end{center}
\caption{
Summary of simulated data sets for this study.}
\end{table}

Seven independent sources of systematic error were taken into account
(see~\cite{mba:hwse} for details)
giving a 
total systematic error of $\sim 1$--5\% over most of the kinematic range
which is a factor of 2--5 better than presently achieved.
Residual systematic effects were
represented by a point to point uncorrelated systematic error of 1\%.

The following model was fitted to the simulated data:
\begin{equation} \label{mba:sysdef}
F_i(p,s) = F_i^{QCD}(p)~~ ( 1 - \sum_l  s_l \Delta_{li}^{syst})
\end{equation}
where 
$F_i^{QCD}(p)$ is the QCD prediction for $F_2$,
$\Delta_{li}^{syst}$ is the (relative) systematic error on
datapoint
($i$) stemming from source ($l$)
and $s_l$ are the systematic
parameters. 
It is assumed that these parameters
are uncorrelated and gaussian distributed with zero mean and unit variance.

The parameters ($p$) in \Gl{mba:sysdef} represent $\as (M_Z^2)$ and
those parameters describing the parton distributions at the input scale 
$Q^2_0 = 4$~\gev.
The gluon distribution
($xg$), the quark singlet distribution ($x\Si$) and the difference
of the up and down quark distributions ($x\De_{ud}$) were
parametrised as
\beq{mba:analeq3}
xh(x,Q_0^2)  =  A_h x^{B_h}(1-x)^{C_h} P(x)
\eeq
with $P(x) = 1$ for $xg$ and $x\De_{ud}$ and
$P(x) = 1 + D\sqrt{x} + E x$ for $x\Si$.
In the studies presented below two types of fit were considered:
(i)~leave the parameters ($p$) and $s_l$ free in the fit and
(ii)~fix the systematic parameters $s_l$ to zero. In the latter fits
the systematic errors 
($\De s_l = 1$) are propagated to the covariance matrix of the fitted
parameters $(p)$ using the technique described in~\cite{mba:pascaud}.
Since we are only interested in the errors the data
were replaced by the model so that the fits immediately
converged.

The results for the total error on $\as$ are given in \Ta{mba:tab2}.
A fit of the proton datasets I and II with a $Q^2$ cut of
of~3~\gev\ 
yields $\De \as (M_z^2) = 0.006\ (0.012)$ depending on whether the
systematic parameters are fitted or fixed (fit~1 in \Ta{mba:tab2}). 
The error on $\as$ is much improved when pertubative QCD is assumed to
be valid at lower values of $Q^2$ and the cut is lowered from 3 to 1
\gev\ (fit~2). 
Doubling the luminosity of the high $Q^2$ sample has no
effect which illustrates the fact that the error on $\as$ is
dominated by the experimental systematic errors (fit~4). 
A modest improvement in
the $\as$ error is obtained when lower energy proton data are
included (fit~6).

In the fits 1--6 
$x\De_{ud}(x,Q^2_0)$ was kept fixed to the nominal input value.
The error on this distribution was taken from the QCD analysis
of ref.~\cite{mba:ZEUSQCD} and contributes 0.004 to the
error on $\as (M_Z^2)$ (not included in \Ta{mba:tab2}). 
This error is eliminated if HERA deuteron data are 
included:
the difference of $F_2^p$ and $F_2^d$ constrains $x\De_{ud}$
which can thus be left free in the fit. It is seen from
\Ta{mba:tab2} that
the error on $\as$ is reduced even though the number of fit parameters
has increased (fit~7).

\begin{table}[t]
\label{mba:tab2}
\begin{center}
\begin{tabular}{clcccc}
  Fit & datasets & $Q^2_{cut}$ (\gev)  & $L$(set II) (pb$^{-1}$) & 
syst.\ fitted
& syst.\ fixed \\
\hline    
1 & I, II & 3 & 500  & 0.006 & 0.012 \\
2 & I, II & 1 & 500  & 0.002 & 0.006 \\
3 & I, II & 8 & 500  & 0.007 & 0.012 \\
4 & I, II & 3 & 1000 & 0.006 & 0.012 \\
5 & I, II & 3 & 10   & 0.010 & 0.015 \\
6 & I, II, III, IV & 3 & 500 & 0.005 &  \\
7 & I, II, V, VI & 3 & 500 & 0.004 & 0.009 \\
\hline    
\end{tabular}
\end{center}
\caption{
The error $\De \al_s(M_Z^2)$ for several fits to Hera data alone.}
\end{table}

To investigate 
if HERA can improve the error on $\as$ from fixed
target data~\cite{mba:MARCALAIN} 
fits were performed including
SLAC~\cite{mba:slac}, BCDMS~\cite{mba:bcdms}
and NMC~\cite{mba:nmc} proton and deuteron $F_2$ data.
To remove higher twist effects a cut $W^2 > 10$ \gev\ was
imposed. A fit to SLAC and BCDMS data alone reproduced
the result of ref.~\cite{mba:MARCALAIN}: $\De \as (M_Z^2) = 0.003$.

Including the fixed target data
the error on $\as$ is much less sensitive to the $Q^2$
cuts imposed and the luminosity of the high $Q^2$ HERA sample.
Depending on the cuts and the available luminosity the fits yield
$\De \as (M_Z^2) =$ 0.0015~--~0.0020 when all systematic parameters are
left free and
$\De \as (M_Z^2) =$ 0.0025~--~0.0035 when they are kept fixed.

Of course, contrary to an analysis of high $x$ fixed target data alone,
the QCD fits to HERA structure functions
result in a joint determination
of both $\as$ and $xg$. For instance a fit to the datasets I and II
(fit~1) with all systematic parameters left
free results in a determination of the gluon distribution with
an accuracy of about 3\% at $x = 10^{-4}$ and $Q^2 = 20$ \gev,
as illustrated in \Ab{mba:fig2}.

\begin{figure}[t!] % fig 1
%\centerline{\epsfig{file=max.ps,height=3.5in,width=3.5in}}
\centerline{
\psfig{file=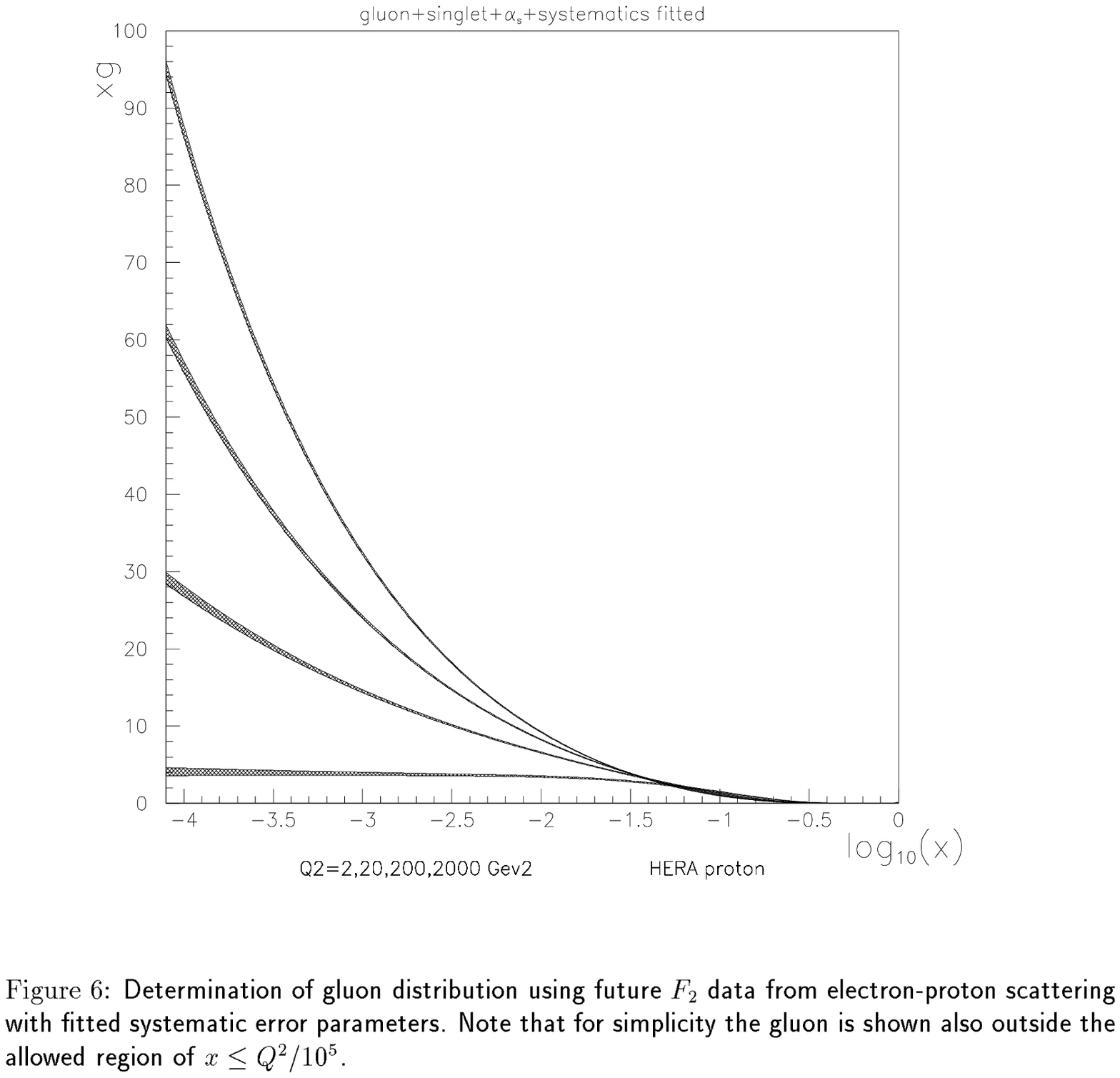,bbllx=80pt,bblly=320pt,bburx=485pt,bbury=727pt,height=9.5cm,clip=}}
\caption{The gluon distribution versus $x$ for different
values of $Q^2$ obtained from a fit to simulated high energy HERA
proton data. The bands show the total error.}
\label{mba:fig2}
\end{figure}

\section*{\centering T\lowercase{heoretical errors on \boldmath{$\alpha_s$}}}

The following possible sources of theoretical uncertainty were
investigated~\cite{mba:hwst,mba:th}:
\begin{itemize}
\item The effect arising from different representations of $\as$.
The scale dependence of the strong coupling constant reads
\beq{mba:eq2}
\frac{\partial a_s(Q^2)}{\partial \ln Q^2} = -\be_0 a_s^2(Q^2)
-\be_1 a_s^3(Q^2) + O(a_s^4)
\eeq
This equation can easily be solved numerically given $a_s$ at
some input scale $Q^2_0$. The following approximate solution
in terms of the QCD scale parameter $\La$ is widely used:
\beq{mba:approx}
a_s(Q^2) = \frac{1}{\be_0 \ln (Q^2/\La^2)}
-\frac{\be_1}{\be_0^3} \frac{\ln \ln (Q^2/\La^2)}
{\ln^2 (Q^2/\La^2)} + O(\ln^{-3}(Q^2/\La^2))
\eeq
Either using \Gl{mba:eq2} or \Gl{mba:approx} causes a shift in
$\as (M_Z^2)$ of 0.001 or less.
\item The offsets originating from the different prescriptions
of the NLO evolution. In the $N$--space approach the evolution
equations and their analytic solutions are usually expanded
as a power series in $a_s$ and terms $O(a_s^3)$ discarded.
The various truncation prescriptions are extensively studied
in ref.~\cite{mba:th}. It turns out that the effect on the 
$Q^2$ evolution of e.g.\ $F_2$ is surprisingly large:
differences of up to 6\% show up at low $x \simeq 10^{-4}$
and are caused by terms in NNLO and beyond. The corresponding
shift in $\as(M_Z^2)$ is estimated to be about 0.003.
\item Renormalisation and factorisation scale uncertainties.
In \Gl{mba:eq1} and \Gl{mba:eq2} the renormalisation scale
($R^2$) and the mass factorisation scale ($M^2$) are both assumed
to be equal to the momentum transfer $Q^2$. The expressions for the
case that $R^2$ and $M^2$ are chosen to be unequal can be found 
in~\cite{mba:th}. \Ta{mba:tab3} gives the shifts in $\as (M^2_Z)$
when these scales are varied independently in the range
$Q^2/4$ to $4Q^2$ for different values of a Q$^2$ cut made on the
data. It is seen that the scale dependence is by far the largest
contribution to the theoretical uncertainty in $\as$ and that
in particular the mass factorisation scale dependence increases
strongly with a decreasing $Q^2$ cut.
\end{itemize}

\begin{table}[t]
\label{mba:tab3}
\begin{center}
\begin{tabular}{ccccc}
  Cut (\gev) & $M^2 = Q^2/4$ & $M^2 = 4Q^2$ & 
               $R^2 = Q^2/4$ & $R^2 = 4Q^2$ \\
\hline    
 4 &   $-$    & $-$0.0120 & $-$0.0076 & +0.0059 \\
20 & +0.0067  & $-$0.0044 & $-$0.0067 & +0.0049 \\
50 & +0.0032  & $-$0.0029 & $-$0.0061 & +0.0042 \\
\hline    
\end{tabular}
\end{center}
\caption{
The theoretical shifts on $\al_s(M_Z^2)$ from scale variations.}
\end{table}

\section*{\centering S\lowercase{ummary}}

An experimental accuracy of $\De \as (M_Z^2) =$ 0.001--0.002
might be in reach provided the following conditions are
satisfied: (i)~$F_2$ measurements become available in the
full HERA kinematic range with systematic and statistical
errors of a few percent only; (ii)~The dependence of the
systematic errors on $x$ and $Q^2$ is sufficiently well known
so that their effects can be absorbed in the QCD analysis
and (iii)~The HERA data can be reliably combined with
fixed target $F_2$ data.

The various prescriptions of the NLO evolution, which differ
by terms of NNLO and beyond, cause a theoretical error
in $\as(M_Z^2)$ of about 0.003. If the renormalisation
scale ($R$) and the mass factorisation scale ($M$)
are varied independently
in the range $Q^2/4$ to $4Q^2$ a theoretical uncertainty
on $\as(M_Z^2)$ of about $\pm 0.005$ ($R$) and $\pm 0.003$ ($M$)
is estimated, provided a $Q^2$ cut of 50~\gev\ is applied. These
errors increase when the $Q^2$ cut is lowered. The theoretical
uncertainties are expected to be reduced substantially once the NNLO
splitting functions become available. 

\section*{\centering A\lowercase{cknowledgements}}

I would like to thank J.~Bl\"{u}mlein, M.~Klein and A.~Vogt
for discussions and comments on the manuscript and
the DIS97
organizing committee for a stimulating
workshop in a beautiful location.

\end{document}